\documentclass{ws-p8-50x6-00}
\begin{document}

\title{Non perturbative QCD}

\author{A. Di Giacomo}

\address{Pisa University and I.N.F.N., Via Buonarroti 2, 56100 Pisa,
Italy\\E-mail: digiaco@mailbox.df.unipi.it} \maketitle \abstracts{
This series of lectures consists of two parts. In the first part
the foundations of perturbative and non perturbative formulation
are discussed. The ambiguity in the definition of vacuum
condensates is then analyzed. In the second part the symmetry
patterns of the deconfining phase transition are discussed.}

\section{Introduction}
I shall preliminary review the basic formulation of quantum field
theory, to clarify what is meant by perturbative and non
perturbative. This will also help to understand the foundations
and the limitations of lattice QCD, which is the main existing non
perturbative approach.
\subsection{Basic Feynman path.}
A quantum system is described by a set of canonical variables
$q(t)$, $p(t)$ and by the Hamiltonian $H$. $q(t),p(t)$ is a short
symbol for $q(i,t)$, $p(i,t)$ the index $i$ running over the
degrees of freedom. When the index $i$ is not specified what we
say will apply to any system, independent of the number of degrees
of freedom. In a quantum field theory $q(i,t)$ are the fields
\begin{equation}
q(i,t) = \varphi_a(\vec x,t)\label{eq:1}\end{equation} the index
being $i\equiv(a,\vec x)$.

The system has a continuous infinity of degrees of freedom.

Solving a field theory means constructing a Hilbert space on which
$q$, $p$ act as operators, obeying the canonical commutation
relations and the equations of motion. A ground state must exist
to ensure stability.

The usual approach consists in splitting the Hamiltonian in a free
Hamiltonian $H_0$ and an interaction $H_I$
\begin{equation}
H = H_0 + H_I \label{eq:2}\end{equation} $H_0$ is quantized and
solved exactly, and the resulting Hilbert space is assumed to
describe also the interacting system. $H_I$ is an operator in that
space and is treated as a perturbation. This approach is called
perturbative quantization.

A more fundamental formulation can be given in terms of Feynman
path integral\cite{1}.

The basic quantity is the action
\begin{equation}
S[q] = \int {\cal L} \,dt\label{eq:3}\end{equation} with ${\cal
L}$ the Lagrangean. $S$ is a functional of the fields. The
equations of motion are the stationary points of the functional
\begin{equation}
\frac{\delta S}{\delta q_i(t)} = 0 \label{eq:4}\end{equation} The
knowledge of the partition function
\begin{equation}
Z = \int\left[ \prod_{i,t} d q(i,t)\right] exp[i S(q)]
\label{eq:1.5}\end{equation} or better of the generating
functional
\begin{equation}
Z[J] = \int\left[ \prod_{i,t} d q(i,t)\right] exp\{i [ S(q) +
\sum_i\int dt J_i(t) q_i(t)]\} \label{eq:6}\end{equation} provides
the solution of the theory.

Indeed the connected time ordered correlators of the fields in the
ground state, are expressed in terms of $Z[J]$ as
\begin{equation}
\langle 0 | T\left( q(i_1,t_1)\ldots q(i_n,t_n)\right)|0\rangle =
\frac{1}{Z}(-i)^n\left.\frac{ \delta^n Z[J]}{ \delta
J(i_1,t_1)\ldots \delta J(i_n,t_n)}\right|_{J=0}
\label{eq:7}\end{equation} In terms of the correlators the Hilbert
space is determined, and the solution of the theory is known
(reconstruction theorem).

$Z$ is a functional integral, with a continuous infinity of
integration variables. By definition it is defined as the limit of
a sequences of ordinary integrals, computed on a discretized space
time, on sets of points which become dense in space time in the
limit.

The above statements are clearly illustrated by the following
argument\cite{1}. Consider the transition amplitudes
\begin{equation}
\langle q',t'|q,t\rangle \label{eq:8}\end{equation} between
eigenstates of the $q$'s at different times. Any physical
amplitude will be known if all the amplitudes (\ref{eq:8}) are
known, the $q$'s being a complete set of commuting operators.

Eq.(\ref{eq:8}) can be rewritten
\begin{equation}
\langle q'| e^{-i H(t'-t)}|q\rangle \label{eq:9}\end{equation}
Putting $\delta = \frac{t'-t}{n+1}$,
\[ e^{-i H(t'-t)} = \left[ e^{-i H \delta}\right]^{n+1}\]
as $n\to \infty$, $\delta\to 0$.

By inserting complete sets of states, we get
\[\langle q',t'|q,t\rangle =
\sum_{q_n\ldots q_1} \langle q'| e^{-i H \delta}| q_n\rangle
\langle q_n|e^{-i H \delta}\ldots e^{-i H
\delta}|q_1\rangle\langle q_1| e^{-i H \delta} | q\rangle\] We
shall consider for the sake of simplicity $H$'s of the form $H =
\frac{p^2}{2} + V(q)$, but what follows can be easily adapted to
momentum dependent interactions. Then, by use of the
Baker-Haussdorf formula
\begin{equation}
e^{-i H \delta} \simeq e^{-i\frac{p^2}{2}\delta}e ^{-i
V(q)}\left(1 + {\cal O}(\delta^2)\right)
\label{eq:10}\end{equation}
\begin{equation}
\langle q_{k+1}| e^{-i H \delta} | q_k\rangle\simeq e^{-i
\left[V(q_k)  -  \frac{(q_{k+1} - q_k)^2}{2 \delta^2}\right]
\delta} \label{eq:11}\end{equation} and
\begin{equation}
\langle q' t'| q t\rangle = \lim_{n\to\infty}\int \prod_{k=1}^n
dq_k\,\exp\left[ i \sum_k {\cal L}(q_k)\right] =
\lim_{n\to\infty}\int \prod_{k=1}^n dq_k\,e^{i S_n(q)}
\label{eq:12}\end{equation} Eq.'s (\ref{eq:11},\ref{eq:12}) allow
to continue analitically to imaginary times (Euclidean space time)
by the Wick rotation $x_0 = i t$. The limit of the Euclidean
amplitude when $x'_0 \to T$, $x_0 \to -T$, and $T\to\infty$ can be
evaluated as follows
\begin{eqnarray}
&&\langle q' T|q -T\rangle_E = \sum_{E_{n_1}, E_{n_2}} \langle q'|
e^{- \frac{T}{2}H}| E_{n_1}\rangle \langle E_{n_1}| e^{- [ S(q) +
J\cdot q]}|  E_{n_2}\rangle \langle E_{n_2}| e^{-
\frac{T}{2}H}|q\rangle\nonumber\\ && = \sum_{E_{n_1}, E_{n_2}}
e^{-\frac{T}{2}E_{n_1}}\psi_{E_{n_1}}(q')
 e^{-\frac{T}{2}E_{n_2}} \psi_{E_{n_2}}(q)
\langle E_{n_1}| e^{- [ S(q) + J\cdot q]}|  E_{n_2}\rangle
\label{eq:13}\end{eqnarray} The external current has been taken to
be zero outside the time interval $(-\frac{T}{2},\frac{T}{2})$,
and the $\psi$'s are the wave functions of the states
$|E_{n}\rangle$ which are eigenstates of energy. As $T\to\infty$
only the ground state survives and, apart from irrelevant
multiplicative constants which cancel in eq.(\ref{eq:7}),
eq.(\ref{eq:13}) defines the euclidean version of $Z[J]$, and
selects the true ground state of the theory. The Lagrangean
defines a field theory if the limit $\delta\to 0$ $(n\to\infty)$
exists (ultraviolet limit), and if the limit $T\to \infty$ exists
(infrared limit).

Note that after Wick rotation $Z$ is in fact the partition
function of a statistical system. The infrared limit exists if the
theory ha a mass gap, or, in the language of statistical mechanics
if the thermodynamical limit exists.
\subsection{Lattice QCD}
Lattice formulation of QCD\cite{2} is a clever construction of
approximants in the sequence of discrete integrals which define
the partition function.

The theory is defined on a square lattice in space-time, and the
building block is the link
\begin{equation}
U_\mu(n) = e^{i a g_0 A_\mu(n)} \qquad (\mu = 1\ldots 4)
\label{eq:14}\end{equation} which is the parallel transport from
the site $n$ of the lattice to the neighboring site in the
direction $\mu$. The action is expressed in terms of the parallel
transport along an elementary square, in the plane $\mu \nu$, the
so called plaquette $\Pi_{\mu\nu}$ as
\begin{equation}
S = \sum_{n,\mu\nu} \beta {\rm Re}\,Tr\left\{1 -
\frac{1}{N_c}\Pi_{\mu\nu} \right\}\qquad \left(\beta = \frac{2
N_c}{g^2}\right) \label{eq:15}\end{equation} In the limit of zero
lattice spacing it  tends to the continuum action
\begin{equation}
S\mathop\simeq_{a\to 0} -\frac{1}{4}\sum G^a_{\mu\nu} G^a_{\mu\nu}
\label{eq:16}\end{equation} A critical point exists at
$\beta\to\infty$ ($g^2\to0$) where the $\beta$ function of the
theory tends to zero by negative values (asymptotic
freedom)\cite{3}, and like in any statistical model QCD is defined
as a field theory when approaching that point.

At sufficiently large $\beta$, by renormalization group arguments
the lattice spacing $a$, measured in physical units, becomes
smaller and smaller
\begin{equation}
a(\beta) \sim \exp(- b_0 \beta) \label{eq:17}\end{equation} $-b_0$
being the first coefficient in the perturbative expansion of the
$\beta$ function.

At sufficiently large $\beta$ the correlation length becomes large
with respect to lattice spacing, and the coarse structure of the
lattice irrelevant: the limit of eq.(\ref{eq:12}) exists.

On the other hand if the lattice is kept large compared to the
correlation length also the thermodynamical limit is reached,
since the theory has a fundamental scale built in.

This argument indicates that $QCD$ exists as a field theory, even
if a formal proof of that statement does not yet exist in the
sense of constructive field theory.

At sufficiently large $\beta$, and large enough lattice a good
approximant is obtained for $Z$, in the sense of the limit
(\ref{eq:12}), and in particular the true ground state is
selected, as shown in eq.(\ref{eq:13}).

The formulation is gauge invariant and needs no gauge fixing and
no Faddeev-Popov ghosts.

It would be nice if analytic solutions could be given of the
lattice regularized theory. Unfortunately nobody has been able up
to now to produce them. Numerical simulations are the usual
technique to compute physics from lattice. Euclidean QCD is a
statistical mechanics, and like in statistical mechanics
Montecarlo techniques can be used to compute the regularized
Feynman integrals and correlation functions.

\subsection{Perturbative expansion.}
The action $S$ can be split in two parts: a term $S_0$ which is
bilinear in the fields and a term $S_I$ which includes the rest
\begin{equation}
S = S_0 + S_I \label{eq:b1}\end{equation} $S_0$ can be written as
\begin{equation}
S_0 = \frac{1}{2}\sum_{\alpha\beta} q_\alpha {\cal
D}^{-1}_{\alpha\beta} q_\beta \label{eq:b2}\end{equation} where
$\alpha = i,t$ spans the degrees of freedom and time. Then
$\exp[-S]$ is expanded in powers of $S_I$ as
\begin{equation}
\exp[-S] = \exp[-S_0] \sum_{n=0}^\infty\frac{ (-S_I)^n}{n!}
\label{eq:b3}\end{equation} and
\begin{equation}
Z = \int\prod_\alpha dq_\alpha \exp[-S] = \int\prod_\alpha
dq_\alpha e^{-S_0}  \sum_{n=0}^\infty\frac{ (-S_I)^n}{n!}
\label{eq:b4}\end{equation} $S_I$ is in general a polynomial of
the fields and by inserting eq.(\ref{eq:b3})
\begin{equation}
\langle T(q_{\alpha_1}\ldots q_{\alpha_n}\rangle = \frac{1}{Z}
\int\prod_\rho dq_\rho \exp(-\frac{1}{2} q_\alpha {\cal
D}^{-1}_{\alpha\beta} q_\beta) q_{\alpha_1}\ldots q_{\alpha_n}
\sum_{n=0}^\infty\frac{ (-S_I)^n}{n!} \label{eq:b5}\end{equation}
All terms of the expansion are gaussian integrals of the form
\begin{equation}
\frac{1}{Z} \int\prod_\rho dq_\rho \exp(-\frac{1}{2} q_\alpha
{\cal D}^{-1}_{\alpha\beta} q_\beta) q_{\sigma_1}\ldots
q_{\sigma_n} \label{eq:b6}\end{equation} which can be computed and
give
\begin{equation}
\sum_{pairs} {\cal D}_{\alpha_{i_1} \alpha_{i_2}} {\cal
D}_{\alpha_{i_3} \alpha_{i_4}} {\cal D}_{\alpha_{i_{n-1}}
\alpha_{i_n}} \label{eq:b7}\end{equation} where the sum runs over
all possible pairings of indices. Eq.(\ref{eq:b7}) is nothing but
Wick's theorem with ${\cal D}$ the propagator.

In QCD such scheme of quantization produces an $S$ matrix
describing scattering of quarks and gluons, and the ground state
is Fock vacuum. The theory can be renormalized producing for any
observable quantity a power series  of the renormalized coupling
constant $\alpha_s$, with finite coefficients
\begin{equation}
{\cal O} = \sum {\cal O}_r \alpha_s^r \label{eq:b8}\end{equation}
However the series is not convergent, not even as asymptotic
series. This reflects an instability, probably due to the
inadequacy of Fock vacuum as ground state. It is phsically obvious
that the theory describing scattering of particles which are never
observed cannot be a good starting point for a perturbative
expansion\cite{4,13}.

It is an empirical and not fully understood fact, however, that a
few terms of the expansion (\ref{eq:b8}) give a good description
of phenomena involving short distances.

\subsection{Instanton vacuum.}
An attempt to go beyond perturbation theory was developed\cite{5}
after the discovery of classical solutions of QCD with finite
action (instantons), having non trivial topology\cite{6}.

In the computation of an ordinary integral the saddle point
approximation implies a sum over all the stationary points of the
phase:
\begin{eqnarray}
\int dx\,e^{f(x)} &\simeq& \sum_{x_n} \int d(\delta x_n) e^{f(x_n)
+ \frac{f''}{2} \delta x_n^2}\sum_k \frac{(\Delta f_n)^k}{k!}
\label{eq:b9}\\
 \Delta f_n &=& f - f(x_n) - \frac{f''}{2} \delta x_n^2
\nonumber\end{eqnarray} Also in the computation of the Feynman
integral by the saddle point approximation all the stationary
points of the action should be considered.

The usual perturbative expansion consists in retaining only the
trivial stationary point $q=0$, and the saddle point expansion is
nothing but eq.(\ref{eq:b5}). If other minima exist, with finite
action, then, the gaussian approximation reads
\begin{equation}
Z= \sum_n \int d\delta q_n e^{- S^0_n[q]} e^{-\frac{1}{2}\left.
\frac{ \delta S}{\delta q_\alpha \delta q_\beta}\right|{S=S_0}
\delta q^\alpha \delta q^\beta} \sum_k \frac{(\Delta
f_n)^k}{k!}\label{eq:b10}\end{equation} $S^0[q_n]$ is the action
of the classical configuration $q_n$, $\delta q_n = q - q_n$,
$\Delta S = S - S^0[q_n] - \frac{1}{2} \left.\frac{\delta^2
S}{\delta q_\alpha \delta q_\beta}\right|_{q_n} \delta q^n_\alpha
\delta q^n_\beta$.

The existence of minima, i.e. of stable classical solutions, is
due to topology. Finite action requires that, in the euclidean
space time\cite{6},
\[ F_{\mu\nu} F_{\mu\nu} \mathop\simeq_{r\to\infty} {\cal
O}\left(\frac{1}{r^4}\right)\] i.e. that the gauge field $A_\mu$
on the $3d$ sphere at infinity $S_3$ is a pure gauge. This defines
a mapping of $S_3$ into the gauge group. To each continuous
configuration an integer is associated which indicates how many
times the group manifold is swept in this mapping. If $d U$ is the
volume element of the group normalized to 1, ($\int d U = 1$), and
$d \sigma^\mu$ the volume element of $S_3$
\begin{equation}
d U = K^\mu d \sigma^\mu\qquad \int K^\mu d\sigma^\mu = n
\label{eq:b11}\end{equation} By Gauss theorem, if $\partial^\mu
K_\mu = Q(x)$
\begin{equation}
\int Q(x) d^4 x = n \label{eq:b12}\end{equation} $Q(x)$ is known
as topological charge density.

Explicit form of the instanton was produced in ref.\cite{6}.

At the end of the 70's a big effort was done to include non
perturbative effects via eq.(\ref{eq:b10}), i.e. by taking into
account extra terms beyond the trivial stationary point $q=0$. The
hope was that small instantons would dominate and hence a few
terms of the perturbative expansion around them would be a good
description of hadronic physics\cite{5}.

The attempt failed due to difficulties in controlling the infrared
problems, and to the difficulties in explaining confinement. It is
known, however, that instantons play an important role in
understanding chiral properties of the theory\cite{7}.
\subsection{Non perturbative approaches.}
We have no real knowledge of non perturbative methods in field
theory. Only numerical simulations on lattice allow to extract
information on the mechanisms of the theory, e.g. on confinement.

A truly non perturbative idea which has proved to be fruitful is
the $N_c\to\infty$ limit\cite{8}.

The idea is that the main properties of gauge theories are
independent of $N_c$, and the same as for $N_c\to\infty$
($1/N_c\to 0$), with $g^2 N_c = \lambda$ fixed. Corrections for
finite $N_c$, are perturbations which do not alter the structure
of the theory.

Fermion loops are of order $N_f/N_c^{n/2}$, ($N_f$ is the number
of light quarks, $n$ the number of vertices). Except for $n=2$,
the graphs which enters in the definition of the $\beta$ function,
all the rest should be a small correction. Lattice simulation do
indeed indicate that, except for a rescaling of the masses due to
the $\beta$ function, numerical determination in quenched
approximation differ in general by few percent $(\sim 10\%$) from
the determinations in full QCD.

The $1/N_c$ limit gives an explanation of the so called $U(1)$
problem, relating it to a quantity, the topological susceptibility
of the vacuum $\chi$\cite{9a}. Lattice determination of $\chi$
confirm the validity of the approach\cite{9b}. The same idea
proves a useful guiding principle in understanding confinement
(see sect.3).

\section{Non perturbative effects in QCD. Vacuum condensates.}
\subsection{S.V.Z. Sum rules.}
Consider the Wilson operator product expansion of the correlator
of two currents
\begin{eqnarray}
\Pi^{\mu\nu}(x) = T\left( j^\mu(x) j^\nu(0) \right) &=&
C_I^{\mu\nu}(x)\,I + C_4^{\mu\nu} \frac{\beta(g)}{g}
G^a_{\alpha\beta} G^a_{\alpha\beta} \nonumber\\ &&+
C_\psi^{\mu\nu}(x) m \bar\psi\psi +\ldots
\label{eq:c1}\end{eqnarray} In principle this expansion is  a
theorem in perturbation theory\cite{10}, but is presumably valid
in QCD at short distances\cite{11}. In Fourier transform and
renormalizing
\begin{eqnarray}
\Pi_{\mu\nu}(q) &\equiv& \int d^4x\, e^{i  qx }\, \langle 0 |
T(j_\mu(x) j_\nu(0)|0\rangle =\nonumber\\
 &=&(q_\mu q_\nu - q^2
g_{\mu\nu})\left[ \Pi(q^2) - \Pi(0)\right] \label{eq:c2}\\
\Pi(q^2) - \Pi(0) &=& C_1 + \frac{C_{4}}{q^2} G_2 +
\frac{C_\psi}{q^4} G_\psi +\ldots \label{eq:c3}\end{eqnarray}
$
G_2 = \langle 0 | \frac{\beta(g)}{g} G^a_{\mu\nu}
G^a_{\mu\nu}|0\rangle
$ 
has dimension $[m^4]$, is the vacuum expectation value of the
dilatation anomaly and is known as gluon condensate.

$G_\psi = \langle0|m \bar\psi\psi|0\rangle$ also has dimension
$[m^4]$, is related to the breaking of the chiral symmetry and is
known as the quark condensate.

$C_1, C_4, C_\psi$ are dimensionless constants, apart from
$log$'s, and depend on $\alpha_s$. $G_2, G_\psi$ are not defined
in perturbation theory. A dispersion representation for
$\tilde\Pi(q^2) = \Pi(q^2) - \Pi(0)$ is
\begin{equation}
\tilde\Pi(q^2) = - q^2\int_{4 m_\pi^2}^\infty \frac{d s}{ s}
\frac{ R(s)} {s - q^2 + i\varepsilon} \label{eq:c4}\end{equation}
with $R(s)$ the well known ratio
\begin{equation}
R(s) = \frac{\displaystyle \sigma_{e^+ e^-\to h}(s)}{
\displaystyle \sigma_{e^+ e^-\to \mu^+\mu^-}(s)}
\label{eq:c5}\end{equation} In perturbation theory, at each finite
order only $C_1$ gets contributions and is a constant modulo
$log$s. $\sigma\sim 1/s$, modulo $log$s.

Physics, however, is not scale invariant: resonances break scale
invariance by powers, and condensates should describe that
breaking. This is the basic idea of SVZ sum rules where the
equality ``on average'' is exploited
\begin{equation}
- q^2\int_{4 m_\pi^2}^\infty \frac{d s}{ s} \frac{ R(s)} {s - q^2
+ i\varepsilon} \simeq
 C_1 + \frac{C_{G_2}}{q^4} G_2 + \frac{C_\psi}{q^4} G_\psi
\label{eq:c6}\end{equation} Equality ``on the average'' means that
the two sides are not equal at each value of $q^2$, but suitable
averages are equal, which emphasize the region where the two
description are expected to be approximately valid\cite{11}. The
$C$ coefficients are computed to some order in perturbation
theory. A successfull phenomenology results out of which $G_2$ and
$\langle m \bar\psi\psi\rangle$ can be extracted.

From a recent review\cite{12} we quote
\begin{eqnarray}
G_2 &=& (0.24\pm 0.011)\,{\rm GeV}^4 \label{eq:c7}\\
\langle\bar\psi\psi\rangle_{1 {\rm GeV}^2} &=& - 0.13 \,{\rm
GeV}^3 \label{eq:c8}\end{eqnarray} Two questions arise
\begin{itemize}
\item[a)] Is the Wilson O.P.E. eq.(\ref{eq:c1}) well defined?
\item[b)] Can the constants $G_2$, $C_\psi$ be determined by a non perturbative
formulation of the theory, like lattice?
\end{itemize}
The answer to the first question is that it is ambiguous, due to
the presence of renormalons\cite{13}. The answer to the second is
yes, modulo the same ambiguity.

In spite of this ambiguity, however, keeping the leading terms of
the perturbative expansion, sum rules do work, and so does the
lattice determination, which agrees with phenomenology.

We will sketch the argument for the first question, we will show
that it applies in the same way to lattice determinations, and
will present anyhow the successful results from lattice, which
agree with the values (\ref{eq:c7},\ref{eq:c8}).

The perturbative expansion of any observable, e.g. the correlator
of two currents like $\Pi_{\mu\nu}$, eq.(\ref{eq:c3}), is finite
order by order in perturbation theory, i.e.
\begin{equation}
C_1(\alpha_s) = \sum_n C_1^{(n)} \alpha_s^n
\label{eq:c9}\end{equation} where $\alpha_s$ is the measured
coupling constant, and $C_1^{(n)}$ are finite numerical
coefficients. It can be shown that\cite{4}
\begin{equation} C_1^{(n)} \propto
n! n^a b^n \label{eq:c10}\end{equation} where $a$ does not depend
on the operator, $b$ does depend on it. Of course the dependence
(\ref{eq:c10}) on $n$ makes the series non convergent. However the
series could be considered as the asymptotic expansion of a
convergent series if the Borel transform
\begin{equation}
\bar C_1 (b) = \sum_{n=0}^\infty  C_1^{(n+1)} \frac{b^n}{n!}
\label{eq:c11}\end{equation} is convergent. In that case, since
\[ \int b^n e^{-b/\alpha_s} db = \alpha_s^{n+1} n!\]
the asymptotic expansion
\begin{equation}
C_1(\alpha_s) - C_1(0) = \int_0^\infty d b e^{-b/\alpha_s} \bar
C_1(b) db \label{eq:c12}\end{equation} is just eq.(\ref{eq:c9}).

Most probably this is not the case for $C_1(\alpha_s)$.

The resummation of a subseries of graphs shows that
$C_1(\alpha_s)$ is ambiguous in principle by terms which mimic the
condensate\cite{14}.

Consider the QCD correction $\Pi_1(q^2)$ to the correlator,
corresponding to the following graphs
\\
\vskip-\baselineskip {\centerline{
\begin{minipage}{\textwidth}
\epsfxsize = 0.85\textwidth {\centerline{\epsfbox{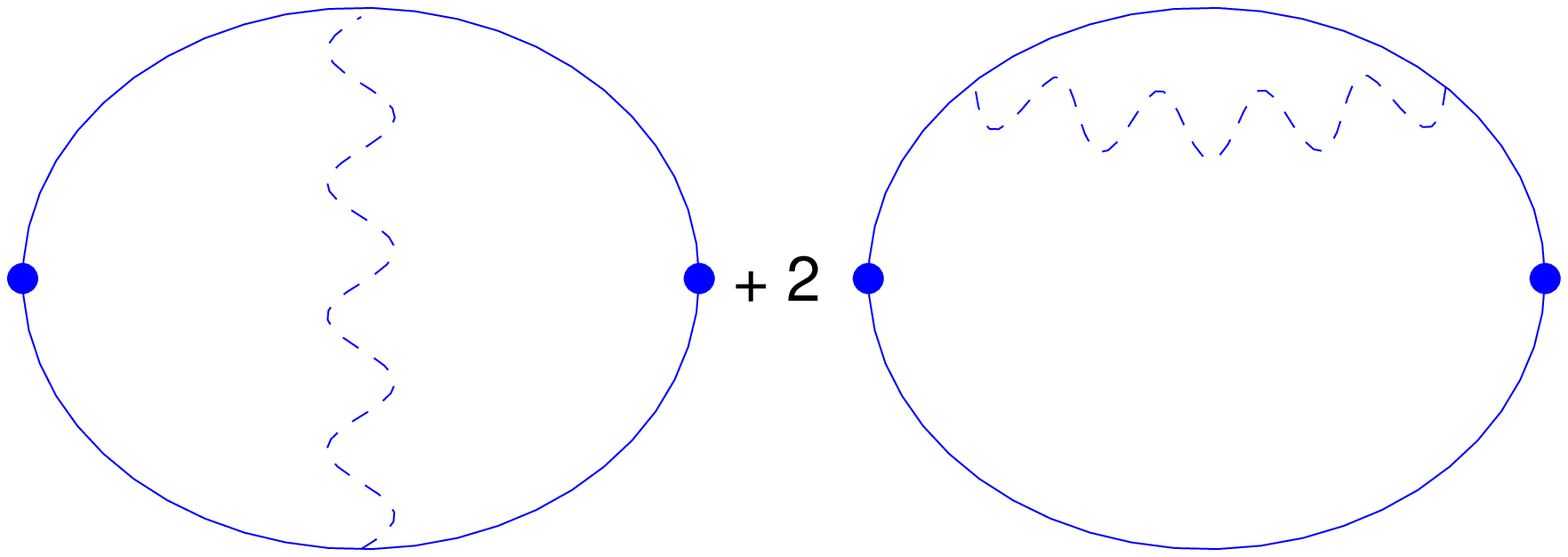}}}
\end{minipage}}}
\vskip0.1in
\begin{equation}
\Pi_1(q^2) = \frac{1}{q^2}\int\frac{d^4 k}{(2\pi)^4} \alpha_s
G(k,q^2)
\end{equation}
Replacing $\alpha_s$ by its running value at one loop
$\alpha_s(k)$, $1/\alpha_s(k) = \frac{1}{\alpha_s} +
\frac{b_0}{2\pi}\ln\frac{k}{\mu}$ amounts to resum a subset of
graphs and gives
\begin{equation}
\Pi_1(q^2) = \frac{1}{q^2}\int\frac{d^4 k}{(2\pi)^4} \left(
\frac{1}{\alpha_s} +\frac{b_0}{2\pi}\ln\frac{k}{\mu}\right)^{-1}
 G(k,q^2) \label{eq:c14}\end{equation}
At small $k$'s, $G(k,q^2)$ is finite and equal to $4 N_c/q^2$ and
hence (\ref{eq:c14}) has a divergence at $k\ll \mu$. The divergent
part is
\begin{equation}
\Pi_1(q^2) \simeq \frac{N_c \alpha_s}{\pi^2 q^4}\int k^3 dk \left(
\frac{1}{\alpha_s}
+\frac{b_0}{2\pi}\ln\frac{k}{\mu}\right)^{-1}\label{eq:c15}\end{equation}
The prescription of the singularity is not defined. Accordingly
there is an ambiguity
\begin{equation}
\Pi_1(q^2) \simeq \frac{N_c \alpha_s}{\pi^2}\frac{\mu^4}{q^4}
e^{-\frac{4\pi}{b_0\alpha_s}} \label{eq:c16}\end{equation} which
mimics the terms proportional to the condensate.

The argument is only an indication, since the subset of graphs
considered is not even gauge invariant. If the indication is
correct it means that the Borel transform is singular, and
condensates cannot be defined

As already anticipated, in spite of that sum rules work and give a
consistent phenomenology.

A similar problem arises when one tries to extract $G_2$ from
lattice\cite{15,16,17}.

Consider the gauge invariant field strength correlator\cite{18}
\begin{equation}
{\cal D}_{\mu\nu,\rho\sigma} = \langle0|Tr\,\left\{ G_{\mu\nu}(x)
S(x,0) G_{\rho\sigma}(0) S^\dagger(x,0)\right\}|0\rangle
\label{eq:c17}\end{equation} where
\begin{eqnarray*}
G_{\mu\nu} &=& \sum_a T^a G^a_{\mu\nu}\\ S_C(x,0) &=& P\exp\left(i
\int_{0,C}^x A_\mu(x) d x^\mu\right)
\end{eqnarray*}
$T^a$ are the generators of the gauge group and $S_C(x,0)$ is the
parallel transport from $0$ to $x$ in the fundamental
representation. $S$ depends on the path $C$: our choice for $C$
will be a straight line.

${\cal D}_{\mu\nu,\mu\nu}$ can be considered as the split point
regulator of $G_2$, \`a la Schwinger.

The OPE gives
\begin{equation}
{\cal D}_{\mu\nu,\mu\nu} \mathop\simeq_{x\to 0} \frac{C_I}{x^4}
\langle I \rangle + C_4  G_2 + \ldots \label{eq:c18}\end{equation}
$<G_2>$ is defined if the term $\propto \frac{1}{x^4}$ can be
unambigously subtracted. Again like for the correlators, higher
order perturbative terms containing log's can sum up and simulate
a constant term, like $G_2$, with an intrinsic ambiguity in the
prescription.

The most general form of the correlator compatible with Poincar\'e
invariance is expressed in terms of two independent
scalars\cite{18} $D(x^2)$, $D_1(x^2)$
\begin{eqnarray}
{\cal D}_{\mu\rho,\nu\sigma} &=& (g_{\mu\nu} g_{\rho\sigma} -
g_{\mu\sigma} g_{\nu\rho})\left[ D(x^2) + D_1(x^2)\right] +
\label{eq:c19}\\ &+& (x_\mu x_\nu g_{\rho\sigma} - x_\mu x_\sigma
g_{\nu\rho} + x_\rho x_\sigma g_{\mu\nu} - x_\nu x_\rho
g_{\mu\sigma}) \frac{\partial D_1}{\partial x^2}\nonumber
\end{eqnarray}
Calling for simplicity $0$ the direction of $x_\mu$ two other
invariants can be defined\cite{16}
\begin{eqnarray}
{\cal D}_{||}(x^2) &=& \frac{1}{3}\sum_{i=1}^3 {\cal D}_{0i,0i} =
{\cal D} + {\cal D}_1 + x^2 \frac{\partial {\cal D}_1}{\partial
x^2} \label{eq:c21}\\ {\cal D}_\perp(x^2) &=&
\frac{1}{3}\sum_{i<j}^3 {\cal D}_{ij,ij} = {\cal D} + {\cal D}_1
\label{eq:c22}\end{eqnarray} ${\cal D}_{||}$ and ${\cal D}_\perp$
can be computed on the lattice\cite{16}.

Their behaviour is shown in fig.1 for quenched theory.
\vskip-\baselineskip {\centerline{
\begin{minipage}{\textwidth}
\epsfxsize = 0.85\textwidth
{\centerline{\epsfbox{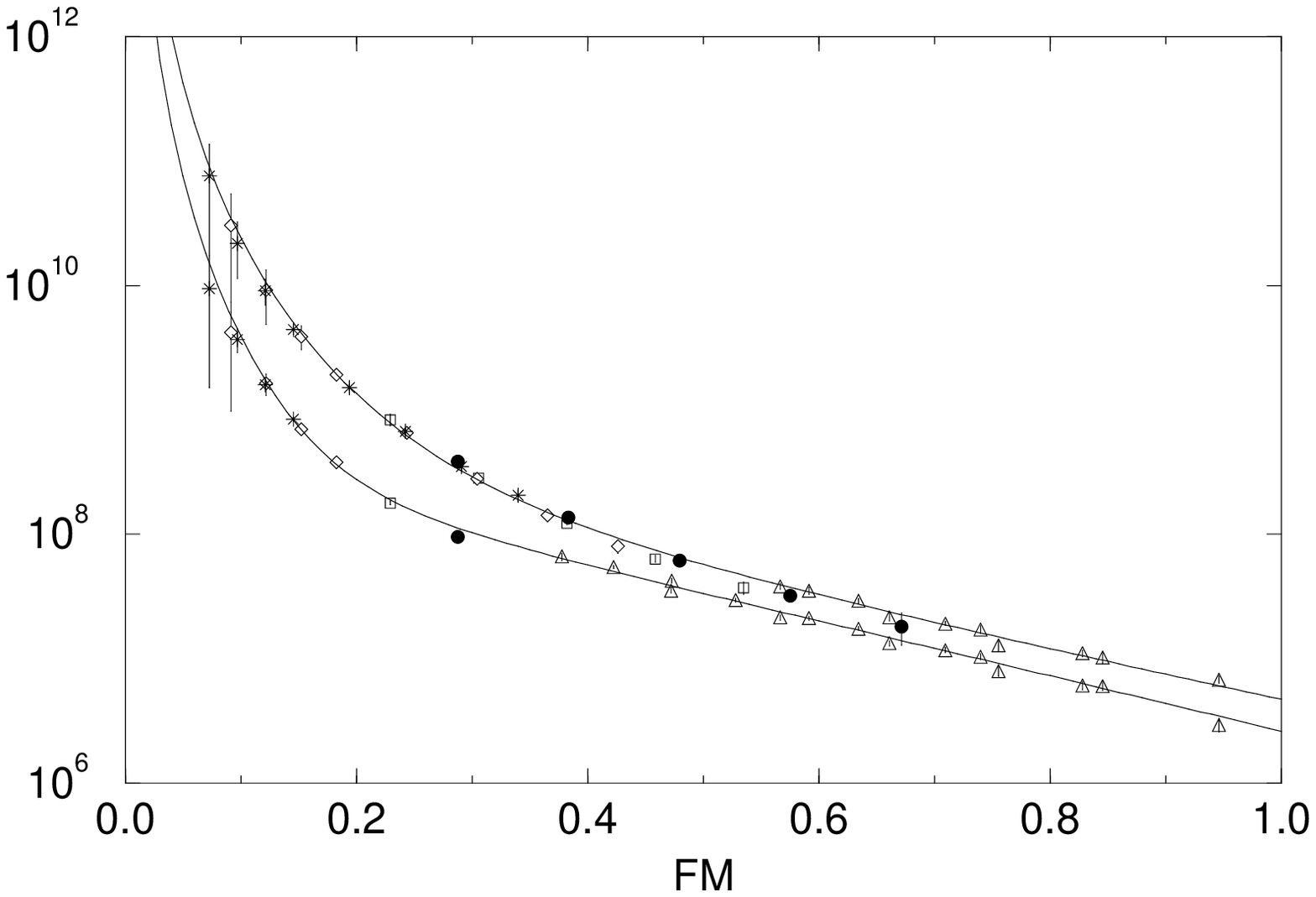}}}
\end{minipage}}}
A parametrization
\begin{equation}
{\cal D}(x^2) = G_2 e^{-\frac{|x|}{\lambda}} + \frac{a}{|x|^4}
e^{-\frac{|x|}{\lambda'}} \label{eq:c23}\end{equation} allows to
extract from the data $G_2$ and $\lambda$.

For quenched $SU(3)$
\begin{equation}
G_2 = (0.15\pm 0.03)\,{\rm GeV}^4\qquad \lambda = (0.22\pm
0.02)\,{\rm fm} \label{eq:c24}\end{equation} For full QCD and 4
quarks at different values of the quark masses (in units of
inverse lattice spacing) the result is smaller than the quenched
one by an order of magnitude and the correlation length is larger
\begin{eqnarray}
am = 0.01 && G_2 = (0.015\pm 0.005)\,{\rm GeV}^4\qquad \lambda =
(0.34\pm 0.04)\,{\rm fm} \label{eq:c25a}\\ am = 0.02 && G_2
(0.031\pm 0.005)\,{\rm GeV}^4\qquad \lambda = (0.29\pm 0.03)\,{\rm
fm} \label{eq:c25b}
\end{eqnarray}
By use of the relation\cite{19}
\begin{equation}
\frac{d G_2}{d m_q} = - \frac{24}{11 - \frac{2}{3}N_f} \langle\bar
q q\rangle \label{eq:c26}\end{equation} and of the  known values
of $\langle \bar q q \rangle$ an extrapolation can be made to the
physical quark mass, giving\cite{17}
\begin{equation}
G_2 = (0.022 \pm 0.005)\,{\rm GeV}^4 \label{eq:c27a}\end{equation}
which compares satisfactoraly with the sum rule determination,
eq.(\ref{eq:c8}).

Old results for $SU(2)$ give\cite{20}
\begin{equation}
G_2 = (0.10\pm0.04)\, {\rm GeV}^4  \label{eq:c27b}\end{equation} A
few words on the technique used in ref.\cite{16,17}. Correlators
measured on a statistical ensemble of configurations are very
noisy due to local quantum fluctuations of the fields. A local
cooling procedure is used to reduce quantum fluctuations\cite{21}.
After a number $n_t$ of cooling steps the size $\lambda$ of the
fluctuations which are cooled away follows a diffusion equation
\begin{equation}
\lambda^2 = D n_t\label{eq:c28}\end{equation} If the distance to
explore is large enough (e.g. 3-4 lattice spacing) cooling can
eliminate local noise and reduce the fluctutions\cite{16,17}.

Field correlators as determined from lattice are useful inputs in
the stocastic approach\cite{18} to QCD vacuum, for which I refer
to the lectures of Y. Simonov of this school.

An alternative procedure to determine $G_2$ is to use the
plaquette as a regulator, and measure the average of the plaquette
itself\cite{15}. A perturbative contribution appears, which is
presumably non Borel summable, which must be somehow subtracted to
isolate the non perturbative part. Again there is ambiguity.
However if the perturbative subtraction is made at some order, the
resulting $G_2$ fits what is found by other methods.

By a similar procedure the quark condensate can be
extracted\cite{17} from the correlator
\begin{equation}
m \langle \bar\psi(x) S(x,0) \psi(0) \rangle\simeq \frac{C_1}{x^2}
+ C_2 \langle m \bar\psi \psi\rangle \label{eq:c30}\end{equation}
The result is, at $\beta = 5.35$
\begin{eqnarray}
|\langle \bar\psi\psi\rangle|^{1/3} = (205\pm .21)\,{\rm
MeV}\qquad am = 0.02 \quad \left(\frac{m_\pi}{m_\rho} =
0.65(3)\right) \label{eq:c31}\\ (160\pm 12)\,{\rm MeV}\qquad am =
0.01 \quad \left(\frac{m_\pi}{m_\rho} = 0.57(2)\right)
\label{eq:c32}
\end{eqnarray}
This approach is like sum rules: it works but it is not understood
why.

\section{Confinement of colour.}
\subsection{Confinement.}
The particles corresponding to the fundamental fields of the
theory, quarks and gluons, have never been observed in nature
(confinement of colour).

In nature the relative abundance of quarks to nucleons has an
upper bound
\begin{equation}
\frac{ n_q}{n_p} \leq 10^{-27} \label{eq:d1}\end{equation} coming
from Millikan like analysis of $\sim 1$~gr of matter.

The expectation in the absence of confinement is\cite{22}
\begin{equation}
\frac{n_q}{n_p} \geq 10^{-12} \label{eq:d2}\end{equation} a factor
$10^{15}$ larger.

A factor $10^{-15}$ cannot be explained by fine tuning of a small
parameter: the only natural explanation is in terms of symmetry.
We shall adopt this prejudice as a starting point.

Lattice provides evidence that confinement is present in QCD. The
well known area law of Wilson loops\cite{23}, $W(R,T) \simeq
\exp(- R T \sigma)$ defines the static potential between $q\,\bar
q$ as
\begin{equation}
V(R) = R \sigma \label{eq:d3}\end{equation} $\sigma$ is known as
string tension: $\sigma\neq 0$ indicates confinement, meaning that
an infinite amount of energy is necessary to pull apart $q$ and
$\bar q$.

Finite temperature QCD can be obtained from the Euclidean
partition function, with a time extension $1/T$, and periodic
boundary conditions for bosons, antiperiodic for fermions.

On the lattice this is done by using a lattice $N_S^3 N_T$ of
extension $N_S\gg N_T$ in each space direction and
\begin{equation}
N_T a = \frac{1}{T} \label{eq:d4}\end{equation} with $a$ the
lattice spacing.

By renormalization group arguments at large enough $\beta$, $a$
depends on $\beta$ as
\begin{equation}
a(\beta) \simeq \frac{1}{\Lambda_L} \exp(-b_0 \beta)
\label{eq:d5}\end{equation} $-b_0$ is the first coefficient in the
expansion of the $\beta$ function, $1/\Lambda_L$ the physical
length scale of the theory.

Because of asymptotic freedom $a$ decreases exponentially to 0 as
$\beta\to\infty$.

The temperature is given by eq.(\ref{eq:d4},\ref{eq:d5}) as
\begin{equation}
T = \frac{\Lambda_L}{N_T} \exp(b_0 \beta)
\label{eq:d6a}\end{equation} and is an increasing function of
$\beta$. At low temperature $\sigma\neq 0$. By increasing $T$ (or
$\beta$) $\sigma$ decreases and at some temperature $T_c$ it goes
to zero. $T_c$ is called the deconfining temperature\cite{25,26}.

An alternative signal for the transition is given by the Polyakov
line
\begin{equation}
{\cal P} = \langle T \oint \exp(i A_0(\vec n,x_0) d x_0)\rangle
\label{eq:d6b}\end{equation} which is the parallel transport along
the time axis from $n_0$ to infinity and back via periodic
boundary conditions.

It can be shown that $P$ is related to the chemical potential of
an isolated quark, $\mu_q$ as
\begin{equation}
{\cal P} = \exp (-\frac{\mu_q}{T}) \label{eq:d7}\end{equation} In
the confined phase $\mu_q$ diverges and ${\cal P} = 0$, while
${\cal P}\neq 0$ in the deconfined phase. Such behaviour is indeed
observed on the lattice, again providing evidence that there is a
deconfining transition\cite{25}.

${\cal P}$ is called an order parameter, being $\neq 0$ in the
weak  coupling regime. $\sigma$ is called a disorder parameter,
being $\neq 0$ in the strong coupling regime.

A drawback of $\sigma$ and $P$  as order parameters is that, in
the presence of quarks they lose their meaning. Indeed the string
should break at some distance creating  $q\bar q$ pairs, and
$\sigma$ is not defined. As for ${\cal P}$, it is an element of
the centre of the group $Z_3$ and signals the breaking of $Z_3$
symmetry in the confined phase. In the presence of quarks $Z_3$ is
not a symmetry. The order parameter usually assumed in full QCD is
the chiral symmetry breaking order parameter $\langle\bar\psi
\psi\rangle$. A priori, however, chiral symmetry is not related to
confinement, although arguments show that it is if the philosophy
of $N_c\to \infty$ is correct. In the same philosophy, however one
should be able to define an order (or disorder) parameter which is
independent of the presence of quarks, at least for not a too
large number of light quarks, which would spoil asymptotic
freedom.

Because of asymptotic freedom the confined phase corresponds to
disorder, the deconfined phase to order.
 In the language of statistical mechanics one should then look for the
symmetry of the system dual to QCD\cite{27}.

\subsection{Order disorder duality.}
Order disorder duality is a property of $d+1$ dimensional systems,
with $d$-dimensional extended configurations $\mu$ having non
trivial topology, and carrying a conserved topological charge.

In the weak coupling regime the system is usually described in
terms of local fields $\Phi$, and the symmetry of the ground state
by order parameters $\langle \Phi\rangle$, which are vacuum
expectation values of some field $\Phi$. A phase transition can
occur at higher couplings, where $\langle \Phi\rangle \to 0$ and
the system becomes disordered.

However a dual description can be given of the system, in which
the configurations $\mu$ become local fields, the original
$\Phi$'s become extended objects, the coupling $g\to 1/g$, and
hence the disordered phase looks ordered, with order parameter
$\langle\mu\rangle$. $\langle\mu\rangle$, the order parameter of
the dual phase is called a disorder parameter (which means order
parameter of the dual).

The prototype example of duality is the $1+1$ dimensional Ising
model\cite{28}. The model is described by a field $\sigma$ which
takes values $\pm1$ on the sites of a 2d square lattice, and is
exactly solvable in the thermodynamical limit. The interaction is
\begin{equation}
- \sum_{k=1}^{2} \sum_i \sigma(i) \sigma(i+\hat k)\qquad Z =
\exp\left\{\beta \sum_{k=1}^{2} \sum_i \sigma(i) \sigma(i+\hat k)
\right\} \label{eq:cc1}\end{equation} The model has 1~dimensional
configurations with nontrivial topology, the kinks, the topology
being identified by the values of $\sigma$ at $x = \pm \infty$ at
fixed $t$.

A kink is a highly non local configuration with
\begin{equation}
\sigma = -1 \quad x \leq x_0\qquad \sigma = +1\quad x > x_0
\label{eq:cc2}\end{equation} A dual variable $\sigma^*$ can be
defined which has also values $\pm 1$ on the sites of the dual
lattice, whose sites corresponds to the links of the original one.

At fixed $t$ $\sigma^*$ is $-1$ if the link joins neighbourghs
sites on which $\sigma$ has the same sign, $+1$ otherwise.

In terms of $\sigma^*$ a link looks as $\sigma^* = +1$
 at $x_0$, $\sigma^* = -1$ everywhere else. The
kink becomes a local excitation in the language of $\sigma^*$.

In ref.\cite{28} it was shown that the partition function in terms
of $\sigma^*$ is again an Ising model with the correspondence
\begin{equation}
Z[\sigma,\beta] = Z[\sigma^*,\beta^*] \label{eq:cc4}\end{equation}
and
\begin{equation}
{\rm sinh}2\beta^* = \frac{1}{{\rm sinh}2\beta}
\label{eq:cc5}\end{equation} The mapping to the dual sends
$\beta\to 1/\beta$ i.e. the weak coupling region into strong
coupling region and viceversa. The high temperature phase which
looks disordered, is ordered and $\langle\sigma^*\rangle$ is its
order parameter. The order parameter of the dual phase is called a
disorder parameter.

Other known systems which admit a dual description and a disorder
parameter are
\begin{itemize}
\item[a)] The $(2+1)$d $XY$ model, (liquid $He_4$). Here the field is
$\theta(i)$ an angle, or better $A_\mu = \partial_\mu \theta(i)$.
The topological excitations are $2$d vortices. Topology is given
by their winding number and
\[ \oint \vec A\cdot d\vec x\]
which is an integer if the field $\theta$ is single valued.
Vortices condense in the disordered phase\cite{29}.
\item[b)] The $(2+1)$d Heisenberg model\cite{30}. Here the topological
excitations which condense in the vacuum above the Curie
temperature, when the magnetization goes to zero, are non abelian
vortices, the instantons of the 2 dimensional $O(3)$ $\sigma$
model. Topology is the winding number of the mapping of the 2
dimensional sphere $S_2$ on $SO(3)$.
\item[c)] In $(3+1)$ dimensional gauge theories the topological excitations are
monopoles\cite{27}, whose topology comes from the mapping $S_2$ on
$SU(2)$ or $S_1$ on $U(1)$. Examples are the $(3+1)$ dimensional
$U(1)$ gauge theory\cite{31}, the $(3+1)$ dimensional non abelian
gauge theory, the $(3+1)$ dimensional supersymmetric QCD\cite{32}.
\end{itemize}

To understand disorder one can adopt two strategies.
\begin{itemize}
\item[1)] Perform the transformation to dual. This has been done in
ref.{\cite{28}} for the Ising model, in ref.{\cite{33}} for the
$U(1)$ compact Villain action, in ref.\cite{32} for the SUSY QCD.
\item[2)] Guess the symmetry of the dual phase and write the disorder
parameter $\langle\mu\rangle$ in terms of the original local
fields $\Phi$. This line has been pursued in ref.\cite{34} on the
continuum and by the Pisa group on the
lattice\cite{27,30,31,35,36}.
\end{itemize}
The advantage of strategy n.2, which we shall adopt, is that the
symmetry can be explored without going through the hardly
manageable procedure of constructing the dual.

The general principle is to create by an operator $\mu$,
constructed in terms of the local fields $\Phi$, a dual excitation
carrying non trivial topological charge, and investigate dual
order by directly computing $\langle\mu\rangle$, the disorder
parameter.

This is done by the field theoretical analog of translation
operator,
\begin{equation}
e^{i p a}|x\rangle = |x + a\rangle \label{eq:cc6}\end{equation}
The operator\cite{37}
\begin{equation}
\mu(\vec y,t) = \exp\left[ i \sum_k\int d^3\vec x \Pi_k(\vec
x,t)\,\bar\Phi_k(\vec x,\vec y)\right]
\label{eq:cc7}\end{equation} with $\Pi_i(\vec x,t)$ the conjugate
momentum to $\varphi_i(\vec x,t)$ and $\bar\Phi_i(\vec x,\vec y)$
the field configuration produced by a topological excitation
located at $\vec y$. In the Schr\"odinger representation, where
the fields $\varphi_i(\vec x,t)$ are diagonal,
\begin{equation}
\mu(\vec y,t) | \varphi_i(\vec x,t) \rangle = | \varphi_i(\vec
x,t) + \bar\Phi_i(\vec x,\vec y)\rangle
\label{eq:cc8}\end{equation} A topological configuration
$\bar\Phi$ has been added to the original one.

Care is needed to adapt eq.'s(\ref{eq:cc7},\ref{eq:cc8}) to
compact theories, where the fields cannot be arbitrarily shifted.
Another important point is that $\langle \mu \rangle$ is only
defined in the thermodynamical limit. Indeed by eq.(\ref{eq:cc7}),
if the number of degrees of freedom is finite $\langle \mu
\rangle$ is an analytic function of $\beta$, and cannot vanish
identically in the ordered phase without vanishing identically.
This is only possible when $V\to\infty$ and Lee-Yang singularities
develop in the partition function. All that can be mastered. We
have computed $\langle\mu\rangle$ for a number of models (in the
3d $XY$ model, the Heisenberg magnet, the $(3+1)$d compact $U(1)$)
where the symmetry of the dual is known. We have thus a tool to
explore the deconfining transition in QCD.

As we shall discuss in QCD the precise symmetry of the dual system
is not known. Configurations candidate to condense in the confined
phase are monopoles. It is indeed a long standing
idea\cite{38,39}, that confinement could be produced  by dual
Meissner effect. Dual here is intended in the sense
electric-magnetic, i.e. that the roles of electric and magnetic
fields are interchanged with respect to an ordinary
superconductor. The chromoelectric field acting between a $\bar q
q$ pair is expected, by this mechanism, to be confined by dual
Meissner effect into Abrikosov flux tubes, with energy
proportional to the length.

The string tension $\sigma$ is then the energy of the tube per
unit length.

The disorder parameter would then be the v.e.v. $\langle
\mu\rangle$ of any operator which creates a monopole.
$\langle\mu\rangle\neq 0$ signals that monopoles condense in the
vacuum, or, under very general assumptions, that the system is a
dual superconductor.

The preliminary question is then to identify the monopoles.

\subsection{Monopoles in non abelian gauge theories.}
Let us first consider $SU(2)$ gauge group, for the sake of
simplicity. Let $\vec\Phi(x)\vec \sigma$ be any local operator in
the adjoint representation, in the usual notations.

We can define the unit vector parallel to $\vec \Phi$ as $\hat\Phi
= \vec\Phi/|\vec \Phi|$, everywhere in a configuration, except at
the zeros of $\vec \Phi(x)$.

Let us define\cite{40} the field operator $F_{\mu\nu}$ as
\begin{equation}
F_{\mu\nu} = \hat\Phi\vec G_{\mu\nu} - \frac{1}{g}(
D_\mu\hat\Phi\wedge D_\nu\hat\Phi)\cdot\hat\Phi
\label{eq:cc10}\end{equation} $F_{\mu\nu}$ is a colour singlet,
defined everywhere except at zeros of $\vec\Phi(x)$, and is
invariant under non singular gauge transformations.

Here
\begin{equation}
\vec G_{\mu\nu} = \partial_\mu\vec A_\nu - \partial_\nu\vec A_\mu
+ g \vec A_\mu\wedge A_\nu \label{eq:cc11}\end{equation} is the
gauge field strength and $D_\mu = \partial_\mu - i g \vec
A_\mu\wedge$ is the covariant derivative.

The coefficient in front of the second term is chosen in such a
way that the bilinear terms in $\vec A_\mu$ cancel between the two
terms. Indeed one has
\begin{equation}
F_{\mu\nu} = \hat\Phi\cdot(\partial_\mu \vec A_\nu -
\partial_\nu\vec A_\mu)
-
\frac{1}{g}( \partial_\mu\hat\Phi\wedge
\partial_\nu\hat\Phi)\cdot\hat\Phi \label{eq:cc12}\end{equation}
Define
\begin{equation}
F^*_{\mu\nu} = \frac{1}{2}\varepsilon_{\mu\nu\rho\sigma}
F^{\rho\sigma} \label{eq:cc13}\end{equation} and
\begin{equation}
j^M_\nu = \partial_\mu F^*_{\mu\nu} \label{eq:cc14}\end{equation}
Then $j^M_\nu$ is zero wherever $\hat\Phi$ is defined, i.e.
everywhere except at the zeros of $\vec \Phi$, and is in any case
conserved since $F^*_{\mu\nu}$ is antisymmetric.
\begin{equation}
\partial_\mu j^M_\mu = 0 \label{eq:cc15a}\end{equation}
Eq.(\ref{eq:cc15a}) identifies a conservation law, a magnetic
$U(1)$ symmetry.

One can operate a gauge transformation bringing $\hat \Phi$ along
a fixed axis. By eq.(\ref{eq:cc12}) then, in all nonsingular
points
\begin{equation}
F_{\mu\nu} = \partial_\mu A^3_\nu - \partial_\mu A^3_\nu
\label{eq:cc15b}\end{equation} is an abelian field. That is why
such transformation is called an abelian projection.

It can be shown that at the singularities of $\hat \Phi$ a non
zero magnetic current is present. The regular field is a magnetic
field of a monopole sitting at the singularity. The singularity of
the gauge transformation which performs the abelian projection
generates the Dirac string which ensures flux
conservation\cite{42}.

The abelian projection of the t'Hooft Polyakov classical monopole
configuration\cite{40} does indeed demonstrate that $F_{\mu\nu}$
is the field of a static Dirac monopole.

In general, for an arbitrary choice of $\vec \Phi$, a conserved
magnetic charge can be defined. So a functional infinity of
conserved monopole charges exists.

Which of them condense in the vacuum to produce dual
superconductivity, if any?

There are in literature two different attitudes on the problem.
\begin{itemize}
\item[a)] (t'Hooft)\cite{27} All monopoles, whatever the choice of $\vec \Phi$,
are physically equivalent. All of them should condense in the
confined phase, to produce dual superconductivity of the vacuum,
and go to normal state above $T_c$.
\item[b)] Some monopole are more equal than others. The maximal abelian
projection identifies the relevant monopoles\cite{43}.
\end{itemize}
We have checked the two alternatives on the lattice as explained
in the following sections\cite{36}.

A similar construction works for $SU(3)$, where two different
monopoles species can be defined for each operator $\Phi =
\frac{\lambda^a}{2} \Phi^a$ in the adjoint representation,
carrying two different magnetic charges\cite{36}.
\section{Some details on the disorder parameter $\langle\mu\rangle$.}
The recipe of eq.(\ref{eq:cc7}) amounts to say that
\begin{equation}
\langle\mu\rangle = \frac{ Z[S + \Delta S]}{Z[S]}
\label{eq:cc16}\end{equation} Indeed
\begin{equation}
\langle \mu\rangle = \frac{1}{Z}\int [\prod d \varphi] \exp[-\beta
S - \beta \Delta S] \label{eq:cc17}\end{equation} Let us consider
$U(1)$ for simplicity
\begin{equation}
\exp[-\beta\Delta S] = \exp\left\{-\beta\left[(
\Pi_{0i}(\theta_{0i} + b_i) - \Pi_{0i}(\theta_{0i} \right]\right\}
\qquad \beta = \frac{1}{e^2} \label{eq:cc18}\end{equation} Here
$1/e b_i(\vec x-\vec y)$ is the vector potential describing the
field of a monopole. $\Pi_{0i} = [\cos(\theta_{0i}) - 1]$ is the
plaquette $0,i$.

In the weak coupling $\theta_{0i} = E_i$, the electric field and
\[ \Delta S \simeq -\frac{1}{e^2}[ e  E_i] b_i\]
 gives exactly the definition
(\ref{eq:cc7}) for $\mu$.

A similar definition can be given for QCD\cite{36}.

Instead of computing $\langle\mu\rangle$, which is affected by
wild fluctuations, it proves convenient to compute
\begin{equation}
\rho = \frac{d}{d\beta}\ln \langle\mu\rangle
\label{eq:cc20}\end{equation} From eq.(\ref{eq:cc16})
\begin{equation}
\rho = \langle S\rangle_S - \langle S + \Delta S\rangle_{S +
\Delta S} \label{eq:cc21}\end{equation} In terms of $\rho$
\begin{equation}
\langle \mu \rangle = \exp \left[ \int_0^\beta \rho(\beta') d
\beta'\right] \label{eq:cc22}\end{equation} since by definition
$\langle\mu\rangle_{\beta=0}=1$.

In the thermodynamical limit one expects in finite temperature QCD
that, if monopole condensation is related to confinement,
\begin{eqnarray}
\lim_{V\to\infty} \langle \mu\rangle &=& 0 \qquad \beta  > \beta_c
\label{eq:cc23}\\ \lim_{V\to\infty} \langle \mu\rangle &\neq& 0
\qquad \beta  < \beta_c \label{eq:cc24}\\ \langle \mu \rangle
&\simeq& (\beta_c - \beta)^\delta \qquad \beta\sim \beta_c
\label{eq:cc25}\end{eqnarray} At finite volume the qualitative
behaviour of $\langle\mu\rangle$ is shown in fig.2, the
qualitative behaviour of $\rho(\beta)$ in fig.3.

{\centerline{
\begin{minipage}{\textwidth}
\epsfxsize = 0.80\textwidth {\centerline{\epsfbox{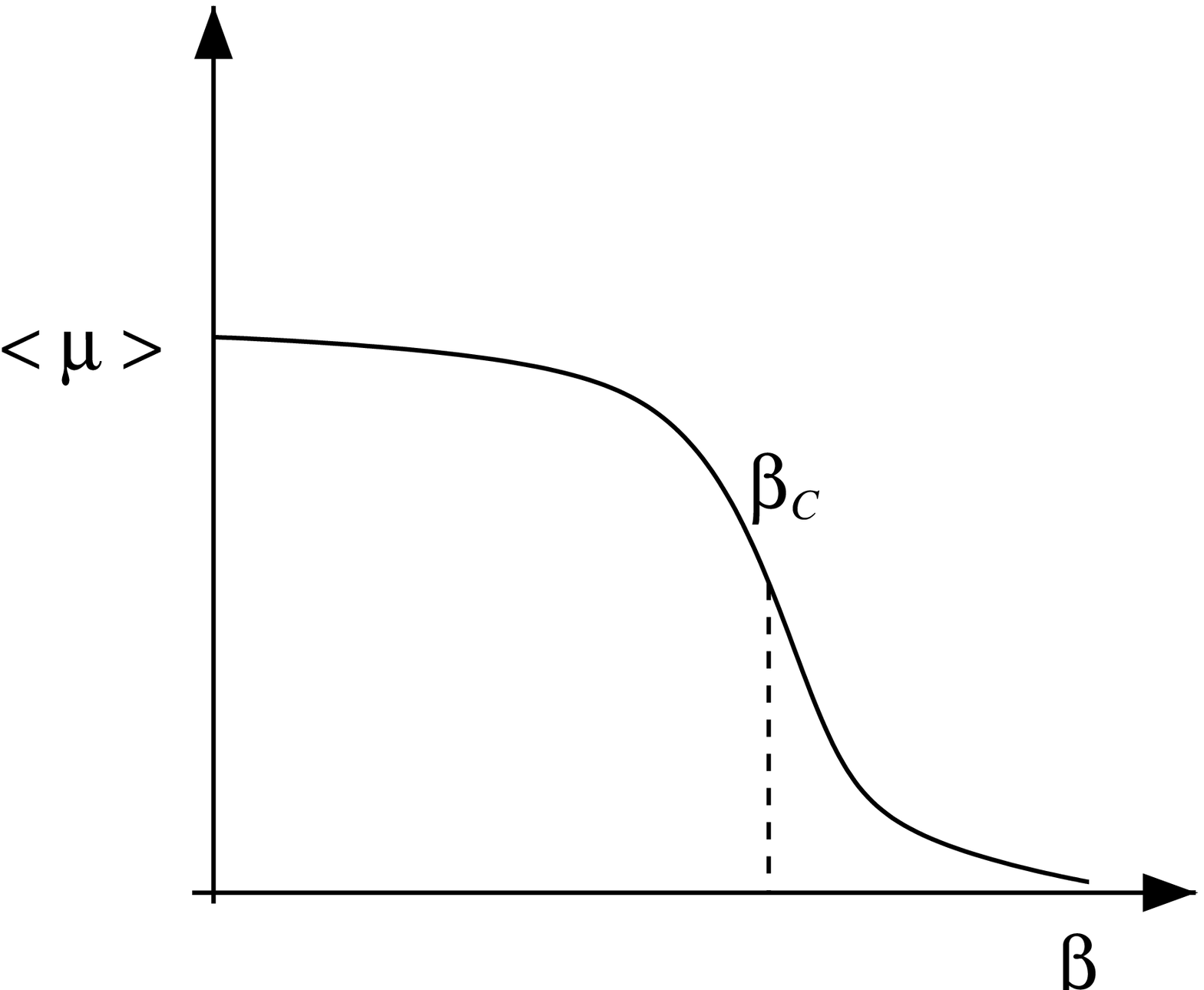}}} Fig.2
\end{minipage}}}

\vskip0.1in \vskip-\baselineskip {\centerline{
\begin{minipage}{\textwidth}
\epsfxsize = 0.8\textwidth {\centerline{\epsfbox{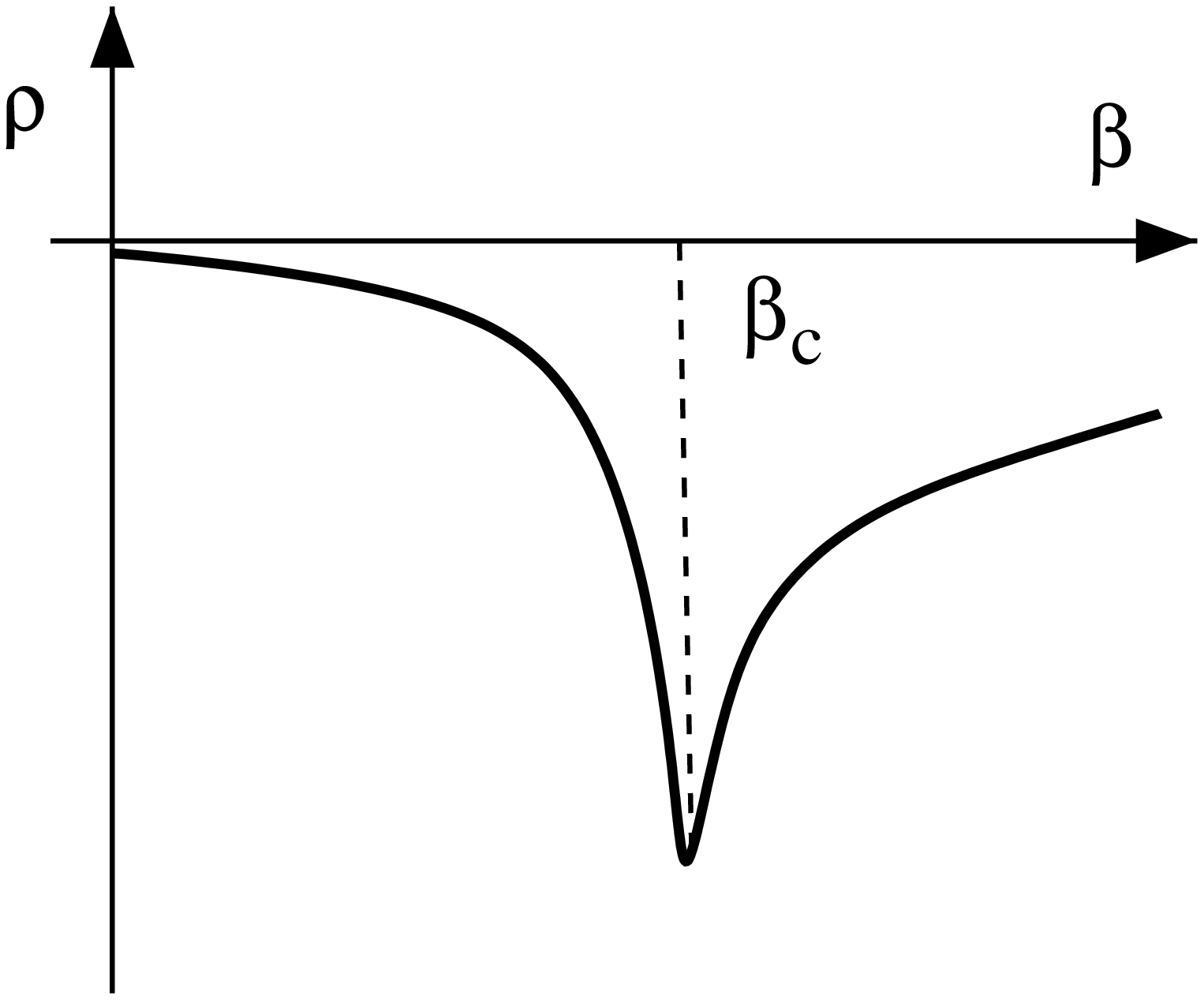}}}
 Fig.3
\end{minipage}}}
\vskip0.1in

For $\beta > \beta_c$ (weak coupling) $\rho$ can be computed in
perturbation getting, in terms of the volume
\begin{equation}
\langle \mu\rangle = \exp( - c_1 V + c_2)\qquad c_1 > 0
\label{eq:cc26}\end{equation} $\langle\mu\rangle$ tends to zero as
$V\to\infty$.

For $\beta < \beta_c$ $\rho$ stays finite at $V\to\infty$, or
$\langle \mu \rangle$ is $\neq 0$.

Around $\beta_c$ the phase transition takes place, which is known
to be second order for $SU(2)$ gauge group, weak first order for
$SU(3)$. This means that the correlation length goes large for
both, since both of them look like second order. On dimensional
grounds
\begin{equation}
\mu \simeq \mu(\frac{\xi}{L},\frac{a}{\xi}) \qquad \xi\simeq
(\beta_c - \beta)^{-\nu} \label{eq:cc27}\end{equation} $\xi$ being
the correlation length and $a$ the lattice spacing.

Near the critical point $a/\xi\sim 0$ and
\begin{equation}
\mu = \mu(\frac{\xi}{L},0) = f(L^{1/\nu}(\beta_c -
\beta))\label{eq:cc28}\end{equation} Eq.(\ref{eq:cc27}) has been
used.

This gives the scaling law for $\rho = \frac{d}{d \rho}\ln \beta$
\begin{equation}
\frac{\rho}{\displaystyle L^{1/\nu}} = F((L^{1/\nu}(\beta_c -
\beta)) \label{eq:cc29}\end{equation} From that scaling $\nu$ and
$\beta_c$ can be determined, together with the critical index
$\delta$.

If condensation is related to confinement the determination should
be consistent with other methods.

The behaviour of $\rho$ for monopoles defined by different choices
of $\Phi$ is shown in fig.4: it looks the same for all of them, in
agreement with the statement of t'Hooft that all monopoles are
physically equivalent.
\vskip-\baselineskip {\centerline{
\begin{minipage}{\textwidth}
\epsfxsize = 0.85\textwidth
{\centerline{\epsfbox{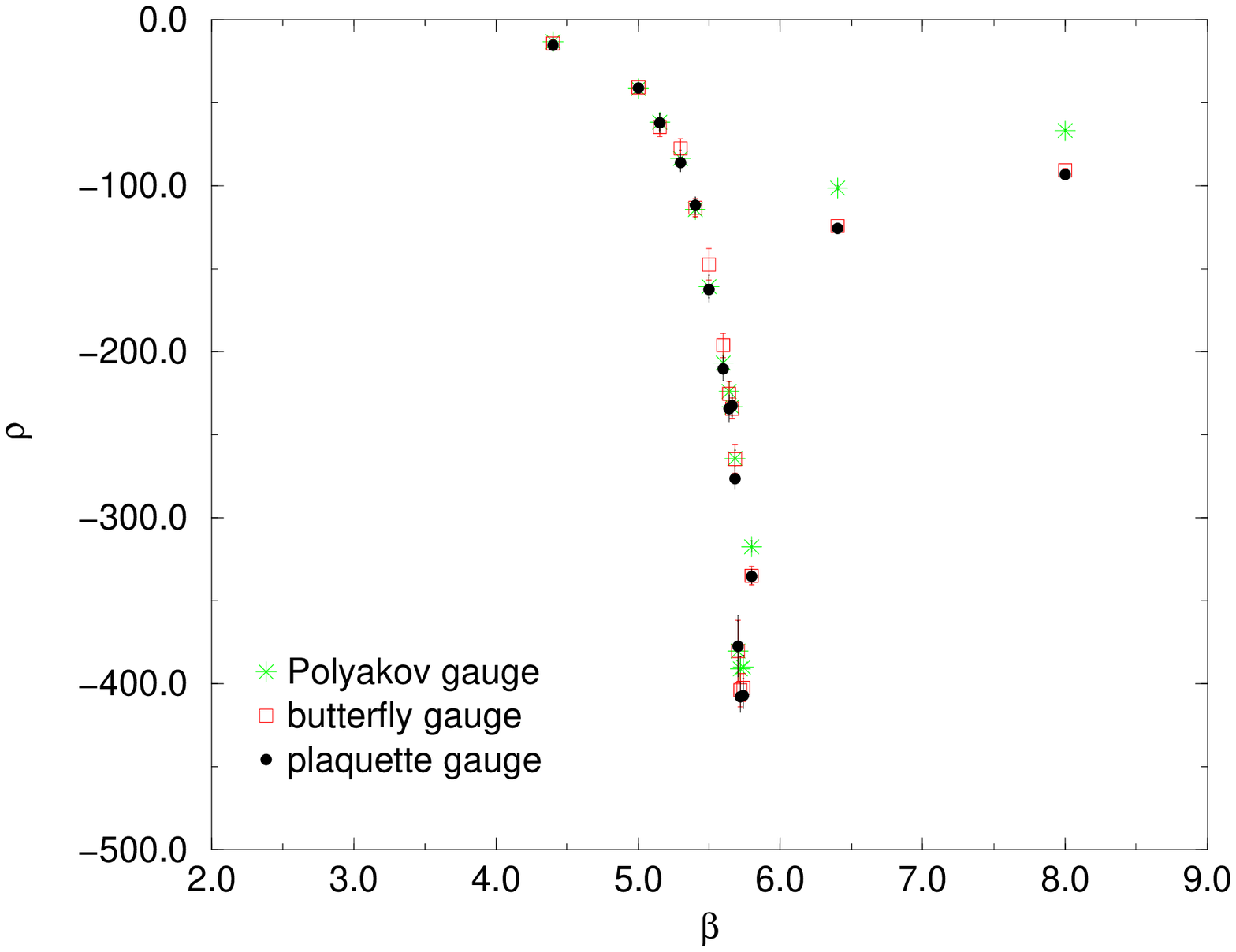}}} Fig.4
\end{minipage}}}
\vskip0.1in

Also the two different monopoles corresponding to the same abelian
projection look identical.

 {\centerline{
\begin{minipage}{\textwidth}
\epsfxsize = 0.75\textwidth {\centerline{\epsfbox{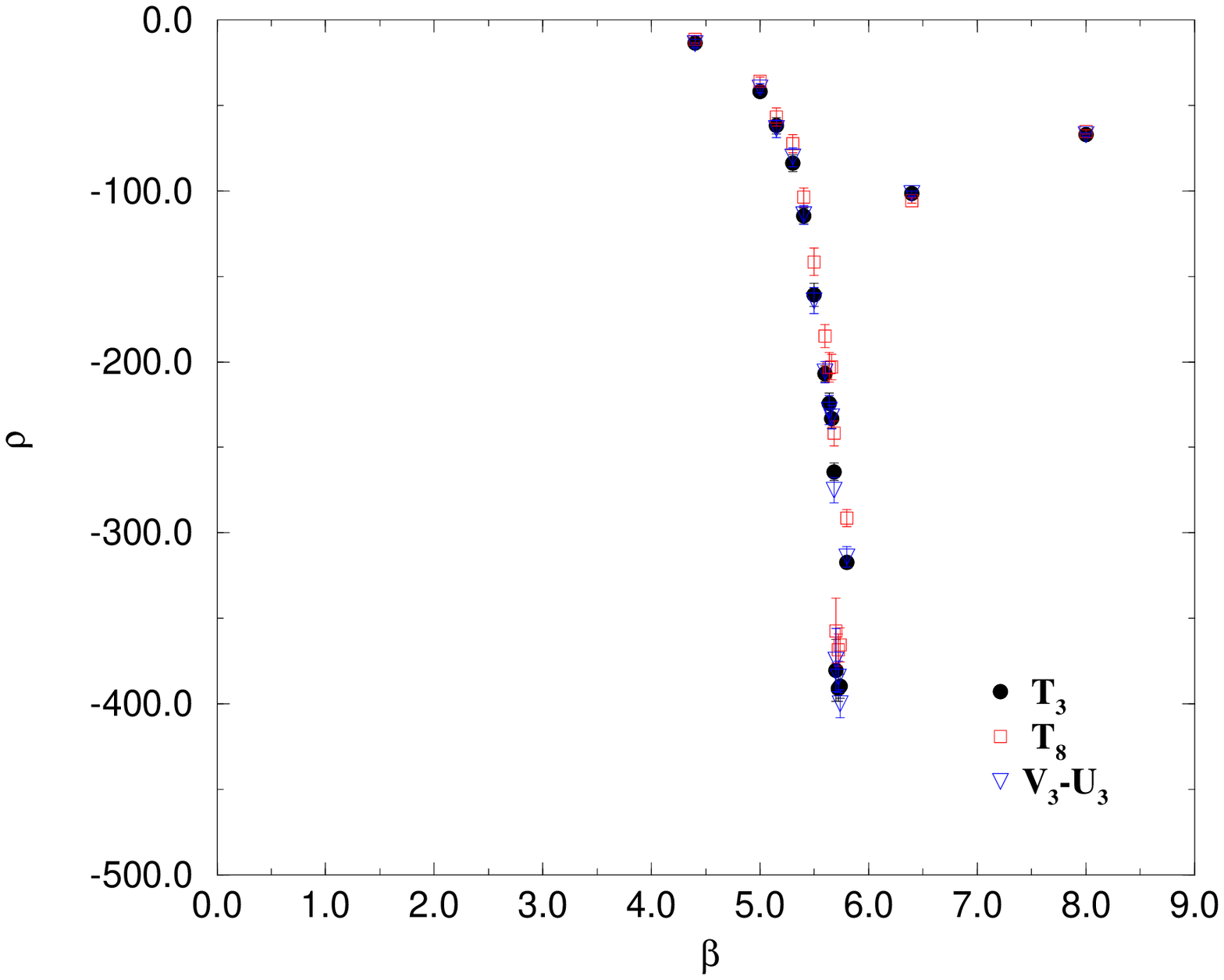}}}
 Fig.5
\end{minipage}}}

{\centerline{
\begin{minipage}{\textwidth}
\epsfxsize = 0.75\textwidth
{\centerline{\epsfbox{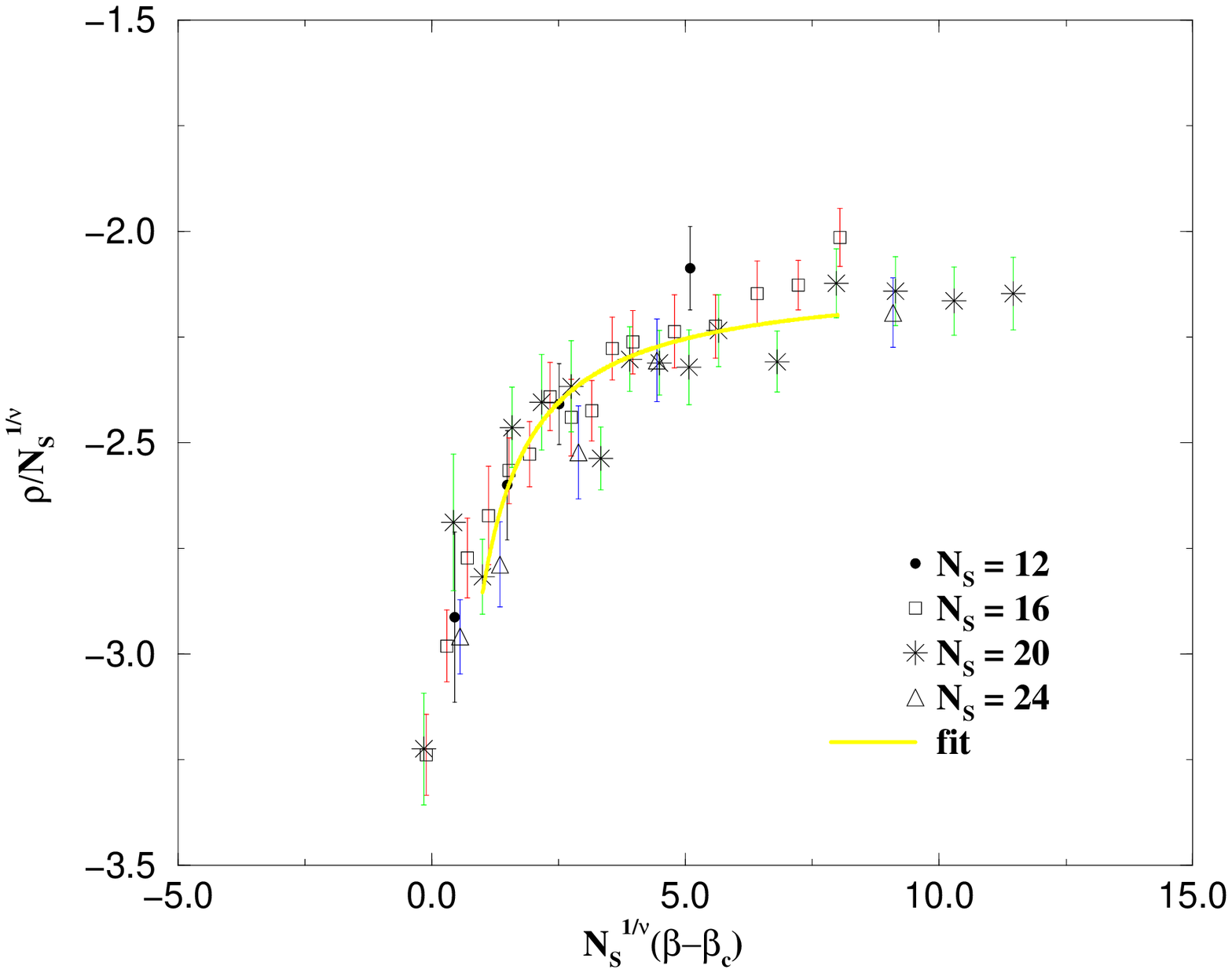}}}
 Fig.6
\end{minipage}}}
\vskip0.1in
 The quality of  scaling law for $SU(2)$ is shown in
fig.6, for the optimal determination of $\beta_c$ and $\nu$.

The result for $\nu$ is
\begin{equation}
\nu = 0.62\pm 0.05\label{eq:cc30}\end{equation} and for $\delta$
\begin{equation}
\delta = 0.18\pm 0.06 \label{eq:cc31}\end{equation} For $SU(3)$
the transition fig.7

\vskip\baselineskip
 {\centerline{
\begin{minipage}{\textwidth}
\epsfxsize = 0.80\textwidth
{\centerline{\epsfbox{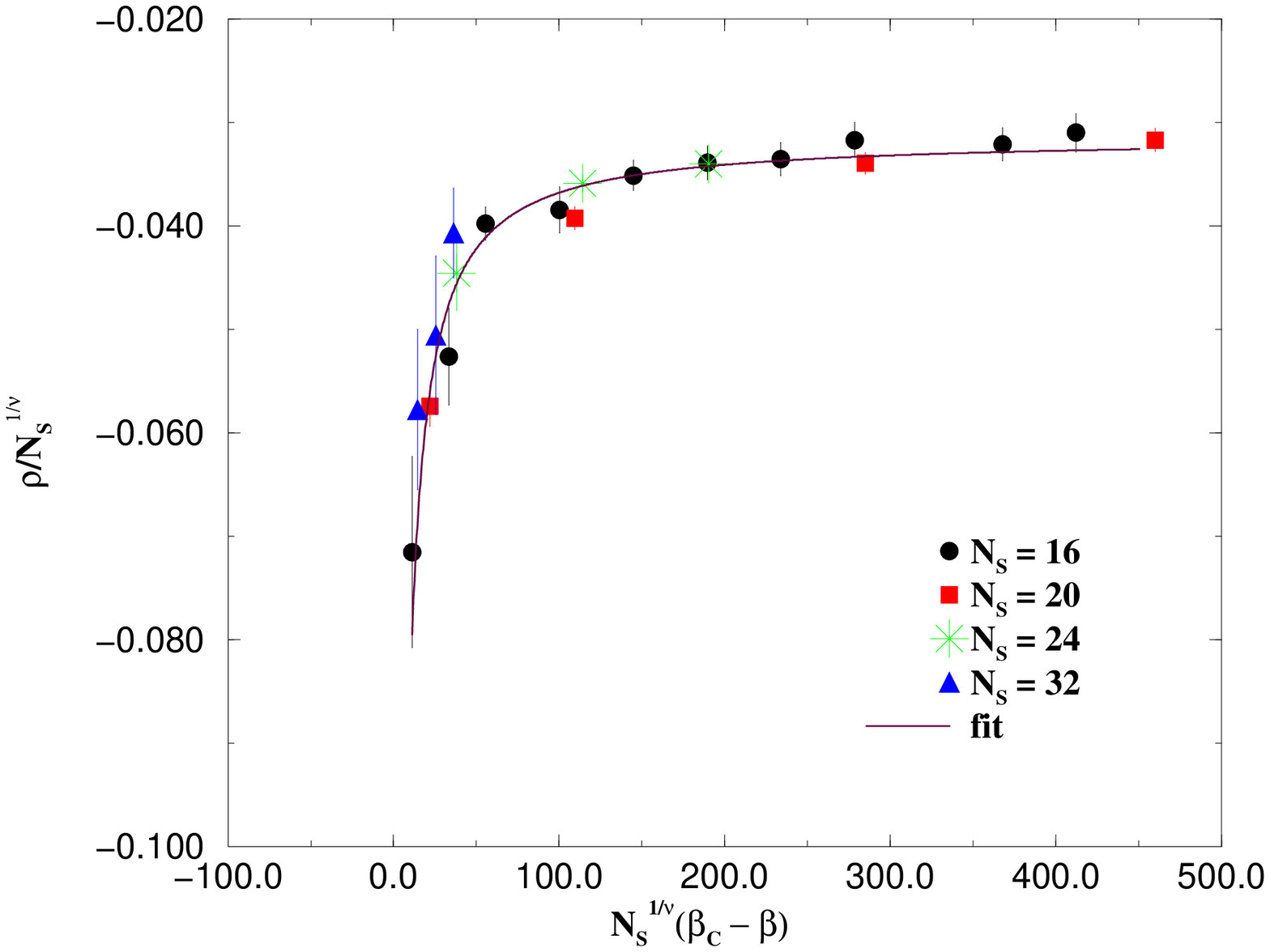}}} Fig.7
\end{minipage}}}
\vskip0.1in

is first order with
\begin{equation}
\nu = 0.33\pm 0.07\qquad \delta = 0.54\pm 0.04
\label{eq:cc32}\end{equation} Monopoles do condense in the vacuum
below $T_c$, supporting thus the mechanism of confinement by dual
superconductivity. Monopole identified by different abelian
projections behave in the same way. As a check the finite size
scaling analysis of $\langle \mu\rangle$ provides determination of
$\beta_c$ and of the critical indices consistent with other
methods.

The disorder parameter $\langle\mu\rangle$ can be defined and
determined also in full QCD, contrary to $\sigma$ and $P$.
Preliminary results give evidence for monopole  condensation also
there.

The method has been checked on 2d Ising model, on 3d $XY$ model,
on Heisenberg magnet, on $U(1)$ with similar results, consistent
with completely different determinations.

\section{The symmetry of the confined phase in QCD.}
What we have learned from lattice is that QCD vacuum is  a dual
superconductor in the confined phase, and goes normal at the
deconfining transition. And this independently of the choice of
the magnetic charge.

This solves the problem that in a given abelian projection some
colored states could be neutral with respect to the abelian
electric charge, and hence not confined.

However the situation is not clear. We do not understand yet the
symmetry pattern of the ground state, except for the fact that any
magnetic charge condensate is a good disorder parameter. There
must exist for sure a more synthetic way of describing this
situation.

An attractive feature of our order parameter is that it is
independent of the presence of quarks, and this  fits the idea
that the $N_c\to\infty$ theory is a good basic description of
physics, ${\cal O}(1/N_c)$ corrections and higher being
perturbations which do not alter its basic features.


\end{document}